%% file: LiverUSRecon.tex
\documentclass[runningheads]{llncs}
\usepackage[T1]{fontenc}
\usepackage{graphicx}
\usepackage{hyperref}
\usepackage{color}

\usepackage{epsfig}
\usepackage{subcaption}
\usepackage{array}
\usepackage{multirow}
\usepackage{adjustbox}
\usepackage{siunitx}
\usepackage{amsmath,amsfonts,mathtools}
\usepackage{booktabs}

\usepackage{pgfplots}
\usepgfplotslibrary{colorbrewer}
\usetikzlibrary{pgfplots.statistics, pgfplots.colorbrewer} 
\usepackage{pgfplotstable}
\usepackage{filecontents}
\usepgfplotslibrary{statistics}
\usetikzlibrary{arrows, arrows.meta}
\usetikzlibrary{calc}

\definecolor{rryellow}{rgb}{0.875, 0.635, 0.294}
\definecolor{rrbrick}{rgb}{0.867, 0.502, 0.169}
\definecolor{rrgreen}{rgb}{0.494, 0.525, 0.216}
\definecolor{rrteal}{rgb}{0.855, 0.871, 0.82}
\definecolor{rrgray}{rgb}{0.631, 0.576, 0.533}
\definecolor{rrlthellow}{rgb}{0.976, 0.965, 0.714}

\usepackage{fancyhdr}
\begin{document}
\title{LiverUSRecon: Automatic 3D Reconstruction and Volumetry of the Liver with a Few Partial Ultrasound Scans}
\titlerunning{LiverUSRecon}

\author{Kaushalya Sivayogaraj\inst{1} \and
Sahan I. T. Guruge\inst{2} \and
Udari A. Liyanage\inst{2} \and 
Jeevani J. Udupihille\inst{3} \and 
Saroj Jayasinghe\inst{2} \and 
Gerard M. X. Fernando\inst{4}\and 
Ranga Rodrigo\inst{1}\and
Rukshani Liyanaarachchi\inst{1}}

%index{Sivayogaraj, Kaushalya}
%index{Guruge, Sahan I. T.}
%index{Liyanage, Udari A.}
%index{Udupihille, Jeevani J.}
%index{Jayasinghe, Saroj}
%index{Fernando, Gerard M. X.}
%index{Rodrigo, Ranga}
%index{Liyanaarachchi, Rukshani}

% \author{Anonymous}
%
% Kaushalya Sivayogaraj, Sahan Guruge, Udari Liyanage,  Jeevani Udupihille, Saroj Jayasinghe, Gerard Fernando, Ranga Rorigo, Rukshani Liyanaarachchi
\authorrunning{K. Sivayogaraj et al.}
% \authorrunning{}
\institute{University of Moratuwa, Moratuwa, Sri Lanka \and
University of Colombo, Colombo, Sri Lanka \and
University of Peradeniya, Peradeniya, Sri Lanka \and
Zone24x7 Inc., USA}

\maketitle              % typeset the header of the contribution
\begin{abstract}
3D reconstruction of the liver for volumetry is important for qualitative analysis and disease diagnosis. Liver volumetry using Ultrasound (US) scans, although advantageous due to less acquisition time and safety, is challenging due to the inherent noisiness in US scans, blurry boundaries, and partial liver visibility. We address these challenges by using the segmentation masks of a few incomplete sagittal-plane US scans of the liver in conjunction with a Statistical Shape Model (SSM) built using a set of CT scans of the liver. We compute the shape parameters needed to warp this canonical SSM to fit the US scans through a parametric regression network. The resulting 3D liver reconstruction is accurate and leads to automatic liver volume calculation. We evaluate the accuracy of the estimated liver volumes with respect to CT segmentation volumes using RMSE. Our volume computation is statistically much closer to the volume estimated using CT scans than the volume computed using Childs' method by radiologists: p-value of $0.094\; (> 0.05)$ says that there is no significant difference between CT segmentation volumes and ours in contrast to Childs' method. We validate our method using investigations (ablation studies) on the US image resolution, the number of CT scans used for SSM, the number of principal components, and the number of input US scans. To the best of our knowledge, this is the first automatic liver volumetry system using a few incomplete US scans given a set of CT scans of livers for SSM. Code and models are available at \url{https://diagnostics4u.github.io/}

\keywords{Liver volumetry \and Ultrasound (US) \and TransUNet \and 3D reconstruction \and Statistical Shape Modeling (SSM).}
\end{abstract}

\section{Introduction}
3D reconstruction of the liver for volume measurement and 3D visual shape analysis using an accessible medical imaging modality like ultrasound imaging is important. It helps clinicians to analyse subject-specific liver morphology and accurately estimate liver volume in real-time. 3D reconstruction from segmentation of 3D scans (slice based 2D image stacks) such as Computed Tomography (CT) and Magnetic Resonance Imaging (MRI) scans, although still demanding, is generally straightforward~\cite{unet,unet++,Vnet,H-DenseUNet,Do-ada,chen2021transunet}. However, the well-known disadvantages of MRI and CT modalities---long acquisition time, cost, and the use of ionizing radiation in CT---make 3D reconstruction using US images attractive. 

3D reconstruction of organs using a few 2D US scans acquired at various angles in different planes is possible~\cite{orex,implicitvol}. However, this technique requires full view of the organ in the scan and uses several input image slices. More crucially, it requires image acquisition location information (pose of the probe) which is difficult to annotate when performing a clinical scan. If liver volume calculation is the only requirement, extracting measurements from left lobe and right lobe of the liver and training a regression model can lead to an estimate of the volumes (called Childs' method by radiologists \cite{childs}). However, measuring lengths from low contrast and noisy US images is subjective, time-consuming, and prone to inter-observer variability; and visual shape analysis is not possible. Moreover, US scans usually do not have the full view of the liver in one image. Thus, 3D reconstruction of the liver using several partial US scans is useful, and current methods are still unable to do so.

3D reconstruction using a few slices where the organ of interest is full in view is not novel. We can examine CT and MRI liver scans as 3D volumes for a qualitative understanding and volumetry using tools such as 3D Slicer~\cite{slicer} and ITK-SNAP~\cite{itksnap}. Reconstructing with CT slices of the left ventricle of the heart has been partially explored by Yuan \emph{et al.}~\cite{yuan2022slice} given an atlas (left ventricle of the heart) and full visibility in CT slices. However, 3D reconstruction of the liver---a large organ in the human body with a complex 3D structure---is challenging due to the partial visibility of the liver resulting from the limited field of view of the US probe, noisiness, and artefacts.
% However, 3D reconstruction of the liver---a large organ in human body with a complex 3D structure---is challenging as, in US slices, the liver is only partially visible in the field of view of the US probe, and US images being affected by noise, artefacts, limited contrast, and limited resolution. The problem is even more challenging in the absence of an atlas.

% 3D reconstruction of organs using a stack of 2D slices is not uncommon for CT or MRI, in contrast to US. 
If the reconstruction must go beyond visualizing CT or MRI volumes, 3D reconstruction from slices is important. There have been approaches that use one or few 2D slices for 3D reconstruction, such as Instantiation-Net~\cite{instantiation-net} for MRI ventricle, liver reconstruction using an X-ray image by Tong~\emph{et al.}~\cite{imagetograph}, liver reconstruction and volume estimation from Topogram image (2D X-ray image taken from CT scanner before going for CT imaging) by Balashova~\emph{et al.}~\cite{balashova20193d} and cardiac 3D reconstruction~\cite{DeepRecon} for MRI. Yuan~\emph{et al.}~\cite{yuan2022slice} use a few 2D CT slices and combine segmentation and 3D reconstruction to reconstruct the left ventricle using a SSM. Tong~\emph{et al.}~\cite{imagetograph} too use an SSM. However, all these methods use X-ray, CT, or MRI images, and the reconstruction is less challenging due to the high contrast, visibility of full extent of the liver and well-defined boundaries as opposed to US.
%where the 3D volume is not available in the form of a stack of slices. 
Therefore, there is no method for reconstructing a large organ like the liver using a few US slices without 3D probe coordinates.
% ImplicitVol~\cite{implicitvol} is a 3D reconstruction method using US slices, e.g., for fatal brains, although it needs the 3D probe locations. 
% To the best of our knowledge, 

%%%%%%%%%%%%%%%%%
In this paper, we create the 3D reconstruction of the liver using just three sagittal plane US slices where the liver is only partially visible with the aid of an SSM. We create the SSM using a population of liver meshes obtained from CT segmentations. The SSM extracts meaningful information and captures the underlying shape variation within the liver population and provides the mean liver model and 
%\textcolor{red}{shape parameters} 
principal components. Using just three slices is advantageous due to the ability to quickly acquire them. A deep network segments the three slices and a Multi-Layer Perceptron (MLP) regressor generates the shape parameters which, in turn, warp the SSM to create a patient-specific 3D reconstruction of the liver. This enables us to accurately estimate the patient-specific liver volume. Our volume estimates are more accurate, i.e., statistically closer to the ground truth (radiologist-segmented CT liver volumes) than the volumes estimated by radiologists using the Childs’ method. To the best of our knowledge, this is the first automated deep learning method that calculates the liver volume from three incomplete 2D US scans. Further, we introduce a new US liver database with parallel, annotated CT scans comprising 134 scans. Our {\bf contributions} are
\begin{itemize}
    \item [--] 3D liver reconstruction and volume estimation using three US scans acquired from mid-line, mid-clavicular line, and anterior auxiliary line of the sagittal plane, where the liver is partially visible,
    \item [--] a database of paired US scans and radiologist-annotated CT scans that comprises 134 such scans, and
    \item [--] surpassing the volume computation accuracy obtained by radiologists using the Childs' method on US images.
\end{itemize}
Our contributions open up an avenue to use less-expensive, noisy, partial US scans of organs for 3D reconstruction and volumetry. This, in our opinion, will make scan-based accurate volume estimation a common place for better diagnosis.

\section{Methodology}
\label{sec:method}
The aim of our framework is to accurately reconstruct the 3D model of the organ that matches with the noisy, possibly partial US scans as few as two or three\footnote{We mention as ``three'' in subsequent discussions for brevity.}. The resulting 3D model is useful for visualization and volumetry in the clinic. There are three main modules in our 3D reconstruction framework: SSM creation, US segmentation, and the 3D reconstruction itself. The SSM module takes a set of manually segmented 3D CT scans of the same organ of multiple subjects after a registration step and produces the mean mesh and principal components. The segmentation module uses TransUNet~\cite{chen2021transunet} to segment the three US images and generates binary masks which guide the final 3D reconstruction module. The 3D reconstruction module is a parametric regression model that warps the average 3D model to match the segmented US images. The average model is the mean of aligned meshes, which has an equal number of vertices and faces as other organ models generated from 3D CT segmentations. The final result is a 3D model of the organ that matches with the three US scans. 

\begin{figure}[t]
        \centering
        \input{files/main_block_diagram}
        \captionsetup{justification=justified}
        \caption{The proposed framework: binary masks of the three US slices generate the shape parameters through the parametric regression MLP. These warp the SSM to generate the 3D liver reconstruction.}
        \label{fig:framework}
\vspace{-1em}        
\end{figure}
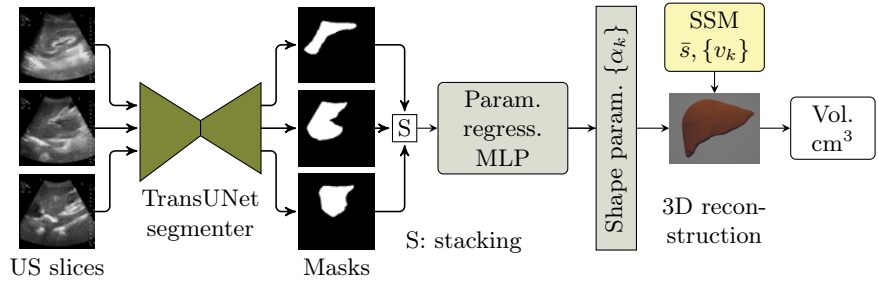

Fig.~\ref{fig:framework} describes this framework, which calculates the liver volume from the reconstructed liver model. 3D reconstruction of the liver is possible by SSM which uses a 3D liver model atlas generated by manually segmented 3D CT scans. Principal Components Analysis (PCA) constructs the parameter space from the generated liver atlas. Raw US slices and their masks train the TransUNet~\cite{chen2021transunet} segmentation network to generate liver masks of the three input US slices. The masks and their shape parameters are the input that train the parametric regression MLP. This MLP, during test time, generates the shape parameters to reconstruct the 3D liver model by warping the SSM. Finally, we calculate the liver volume from the 3D liver mesh.

\noindent {\bf Statistical Shape Model (SSM)}:~The purpose of this model is to produce the 3D liver model that matches the three US scans of a liver. An SSM describes a set of semantically similar objects---3D liver models in our case---using a set of few parameters. It is a fundamental technique in vision and medical image processing invaluable in semantic segmentation and 3D reconstruction \cite{dryden2016}. An SSM has the mean shape of the dataset. This mean shape, combined with principal components that represent the key variations, forms the backbone of the SSM.

We carry out the SSM process introduced in~\cite{gias}. We use a set of $N$ 3D liver meshes generated from CT liver segmentation done by radiologists as the input population. Each liver model has a different number of vertices and faces. As the first step, we carry out a non-rigid registration to fit each 3D liver mesh to the first 3D liver mesh (reference model) to obtain 3D models with the same topology (to make the vertices and faces of each mesh equal in number). Then, we align the fitted meshes rigidly to avoid translational and rotational variations. Then we perform PCA on the generated liver atlas $S = {S_1, S_2,\cdots, S_N}$, $S_i \in \mathbb{R}^{3 \times M}, i \in 1,\cdots, N$, where $M$ is the number of vertices in the reference model, to create the principal components. Each liver shape $S_i$ is mapped to a vector ${s_i}^{T} \in \mathbb{R}^{3M}$. Let $S_{\mathrm{map}} = [s_1, s_2, \cdots, s_N]^T \in \mathbb{R}^{N \times {3M}} $ and the mean mesh
\begin{equation}
\bar{s} = \frac{1}{N}\sum_{i=1}^{N}s_i.   
\end{equation}
Using Singular Value Decomposition (SVD),
$S_{\mathrm{map}} = U\Sigma V^T$, we can represent each liver model using singular vectors $v_k$ (each column of $V$, $V \in \mathbb{R}^{3M \times K}$) as
\begin{equation}
L = \mathrm{reshape}\left(\bar{s} + \sum_{i=k}^{K}v_k\alpha_k\right),
\label{eq:recon}
\end{equation}
where $L \in \mathbb{R}^{3 \times M}$ is a reconstructed liver model. Shape parameters $\alpha_k$ are the variables that represent the liver parametric model. We choose $K=50$ components. In our system, the combination of the segmentation network and parametric regression MLP predicts the shape parameters. We train this network using the three US segmentation masks and ground truth shape parameters. As a result, given a number of US images and CT liver models, and the trained segmentation network and the parametric regression MLP, we can generate the 3D liver model that matches the masks obtained from the three US scans.

\noindent {\bf Segmentation Model for US Liver Segmentation}:~In our 3D liver reconstruction, the binary masks that result from the segmentation of the three US images guide the final 3D reconstruction. We use TransUNet~\cite{chen2021transunet} built based on ResNet50 and ViT~\cite{dosovitskiy2020image} (trained on ImageNet~\cite{imagenet}), and fine-tuned on Synapse multi-organ segmentation dataset and automated cardiac diagnosis challenge dataset~\cite{chen2021transunet}. We fine-tune it using our US liver segmentation dataset. Our dataset comprises three US images each of 134 patients segmented by radiologists. We augmented the dataset using random operations (rotation, translation, flipping, and cropping) when fine-tuning TransUNet. This step prevents overfitting and improves generalization. 
We do not alter any other hyper-parameter of TansUNet. The ViT based TransUNet is important for the segmentation as the partial views of the livers in our US images benefit from the long-range attention available in ViT. In particular, TransUNet adopts a hybrid architectural approach that fuses the strengths of both CNNs and transformers. This hybrid approach combines the fine-grained, high-resolution spatial information inherent in CNN features with the broader global context captured by the transformers. To establish this point, we also used the standard U-Net and different variants of U-Net \cite{unet} to evaluate their performance on this segmentation task in comparison with TransUNet (Table~\ref{ta:accuracy}). In summary, our TransUNet based segmenter accurately segments the noisy, partial US scans of the liver. We feed the segmentation masks to the 3D reconstruction model. 

\noindent {\bf 3D Reconstruction Model for Liver Model Reconstruction}:~Generating shape parameters ($\alpha_k$) from the segmented masks described above is the next step following the segmentation. In this study, we approach the challenge of 3D reconstruction of liver from multiple views of sagittal plane US images. Our objective is to predict the model parameters by directly utilizing the slice-masks as input data. The 3D reconstruction model uses these masks to generate the shape parameters required for 3D liver model reconstruction. The system uses shape parameters ($\alpha_k$s), average liver model ($\bar{s}$), normalized principal components ($v_k$s), and normalization parameters ($E(v_k)$) and $\mathrm{std}(v_k)$) to generate the 3D liver model. To achieve this, we employ a parametric regression MLP that receives the stack of three US slice-masks as its input. So, $\alpha = \mathrm{regression~network}(\mathrm{US~binary~masks})$, where  $\alpha = \{\alpha_1, \dots, \alpha_k, \dots, \alpha_K\}$. Our parametric regression MLP has two layers. 

\noindent {\bf Liver Volume Calculation}:~Following the 3D reconstruction, we are able to estimate the liver volume. We save the 3D reconstruction as an obj file and estimate its volume using trimesh~\cite{trimesh} in $\mathrm{cm}^3$. We have verified the accuracy of trimesh by comparing the volume against the volume computed by 3D Slicer~\cite{slicer}. Our dataset comprises the liver volume estimated by radiologists: 1. using the CT segmentations, and 2. using Childs' method on US slices. We compare the volume we computed using the proposed method with these two methods and statistically analyze.

\section{Experiments and Results}

{\bf Data Acquisition}: We obtained US (three per patient) and corresponding CT scans (for SSM and corresponding 3D reconstruction comparison) of 134 healthy patients\footnote{We plan to make this liver dataset of three annotated US slices and liver annotated CT volumes available for the benefit of the community.}. We captured the three US slices ($1470 \times 2316$) at the mid-line, mid-clavicular line, and anterior auxiliary line of the sagittal plane. An experienced radiologist segmented the liver in US images using ITK-SNAP~\cite{itksnap} to be used for training, and relevant slices of abdomen CT using 3D Slicer~\cite{slicer} to be used for SSM and volume comparison (considered as ground truth). We stacked together the segmented 2D CT slices to reconstruct the 3D liver mesh to be used in the SSM and for comparison. We resized the US images to $192\times 192$ or $384\times 384$ (for ablation). Out of 134 subjects, we allocated 99 for training and 35 for testing. 

\begin{figure}[ht]
  \centering
  \begin{subfigure}[c]{0.18\linewidth}
    \includegraphics[width=\linewidth]{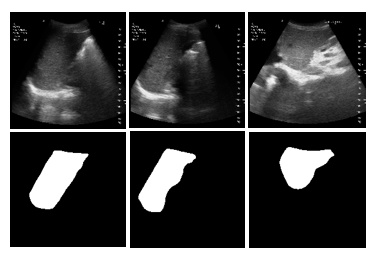}
    %\caption{13\_over}\label{fig:13_over}
  \end{subfigure}
  \begin{subfigure}[c]{0.13\linewidth}
    \includegraphics[width=\linewidth]{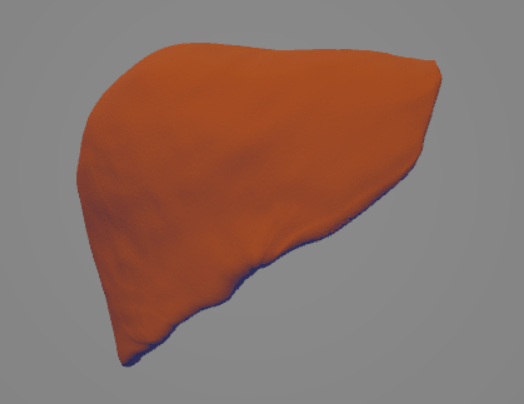}
    %\caption{13\_over\_back}\label{fig:13_over_back}
  \end{subfigure}
  \begin{subfigure}[c]{0.18\linewidth}
    \includegraphics[width=\linewidth]{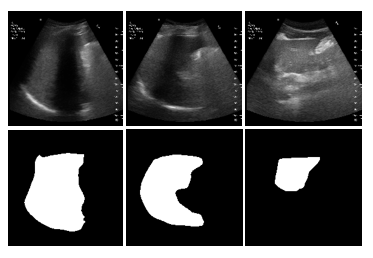}
    %\caption{13\_over}\label{fig:13_over}
  \end{subfigure}
  \begin{subfigure}[c]{0.13\linewidth}
    \includegraphics[width=\linewidth]{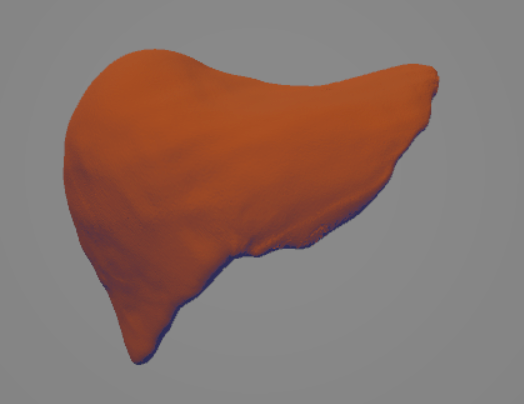}
    %\caption{13\_over\_back}\label{fig:13_over_back}
  \end{subfigure}
  \begin{subfigure}[c]{0.18\linewidth}
    \includegraphics[width=\linewidth]{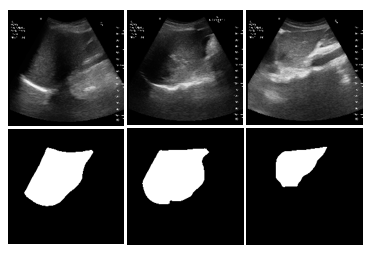}
    %\caption{13\_over}\label{fig:13_over}
  \end{subfigure}
  \begin{subfigure}[c]{0.13\linewidth}
    \includegraphics[width=\linewidth]{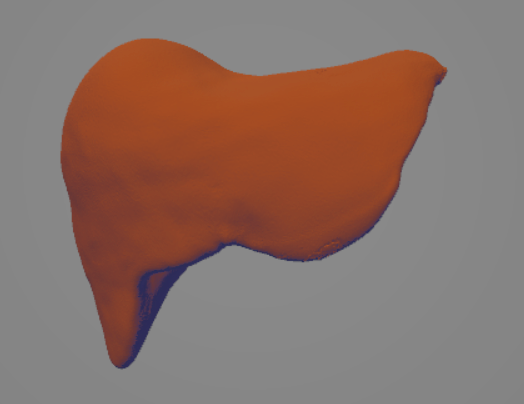}
    %\caption{13\_over\_back}\label{fig:13_over_back}
  \end{subfigure}  

  \begin{tikzpicture}[overlay,remember picture]
    % Adjust the coordinates as needed
    \draw[-Triangle, rrbrick] (-3.8, 0.75) -- ++(0.2, 0);
    \draw[-Triangle, rrbrick] (0.18, 0.75) -- ++(0.2, 0);    
    \draw[-Triangle, rrbrick] (4.15, 0.75) -- ++(0.2, 0);
  \end{tikzpicture}  
  \caption{US segmentation and 3D reconstruction results: Three input  US sagittal plane images, corresponding segmentations, and 3D liver reconstructions using the shape parameters for three subjects.}
  \label{fig:overall_flow_results}
\end{figure}
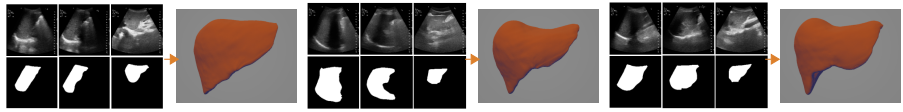

\begin{table}[ht]
\caption{Segmentation accuracy: TransUNet performs better, and hence, was selected for the subsequent experiments. $\ast$ represents the usage of EfficientNet-B7 as an encoder. 3D reconstruction accuracy: CD and MSD are less when we combine TransUNet with Param. Regress. MLP than UNet.}
\label{ta:accuracy}
    \begin{subtable}[t]{0.6\linewidth}
        \centering
        \begin{tabular}{@{}lrrrr@{}}
        \toprule
                   {\bf Segmentation} & FCN   & UNet& UNet++$\ast$ & TransUNet \\ \midrule
        Acc. (\%) $\uparrow$              & 93.2 & 95.4 & 94.4                                                                            & {\bf 97.5}      \\ \hline
        DSC (\%) $\uparrow$  & 38.5 & 65.6 & 68.1                                                                            & {\bf 91.3}      \\ \hline
        HD (mm)    $\downarrow$  & 5.5 & 4.8 & 4.5                                                                            & {\bf 3.6}       \\ \hline
        IoU (\%) $\uparrow$           & 24.1 & 50.2 & 52.7                                                                            & {\bf 84.4}      \\ \bottomrule
        \end{tabular}
    \end{subtable}%
~
    \begin{subtable}[t]{0.4\linewidth}
        \centering
       \begin{tabular}{@{}lrr@{}}
        \toprule
         {\bf Recon.} &TransUNet & UNet \\ 
         {\bf Accuracy}   & + Recon.   & + Recon. \\ \midrule
        MSD (mm)$\downarrow$ & {\bf 6.6} & 6.8 \\ \hline
        CD (mm) $\downarrow$ & {\bf 12.8} & 13.1 \\
        \bottomrule
        \end{tabular}
    \end{subtable}
\end{table}

\begin{figure}[ht]
  \centering
  \begin{subfigure}[b]{0.16\linewidth}
    \includegraphics[width=\linewidth]{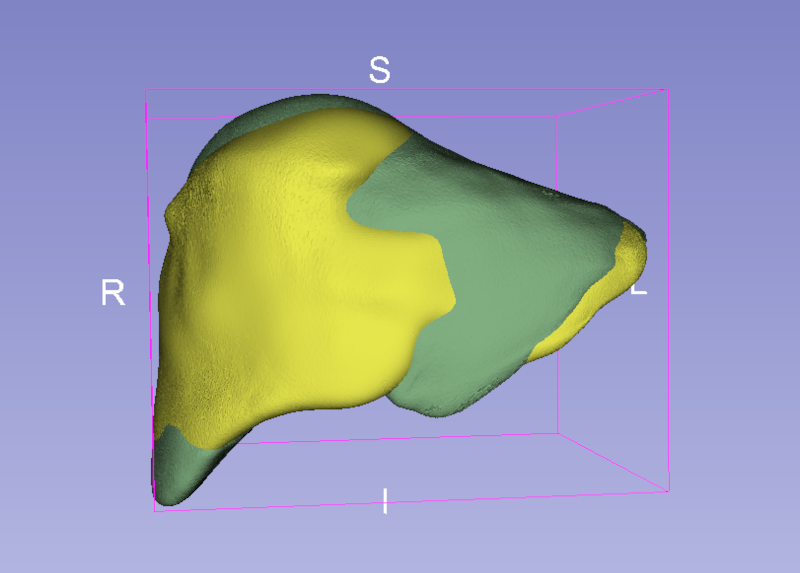}
    %\caption{13\_over}\label{fig:13_over}
  \end{subfigure}
  \begin{subfigure}[b]{0.16\linewidth}
    \includegraphics[width=\linewidth]{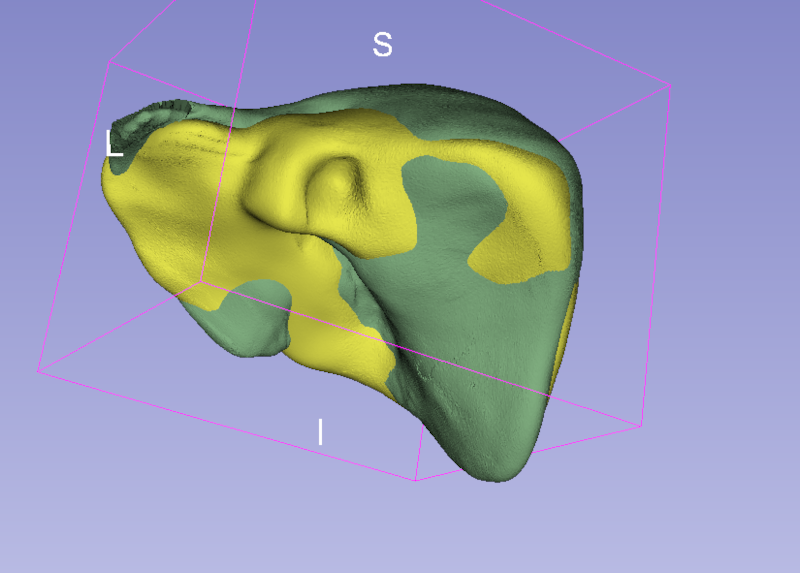}
    %\caption{13\_over\_back}\label{fig:13_over_back}
  \end{subfigure}
  \begin{subfigure}[b]{0.16\linewidth}
    \includegraphics[width=\linewidth]{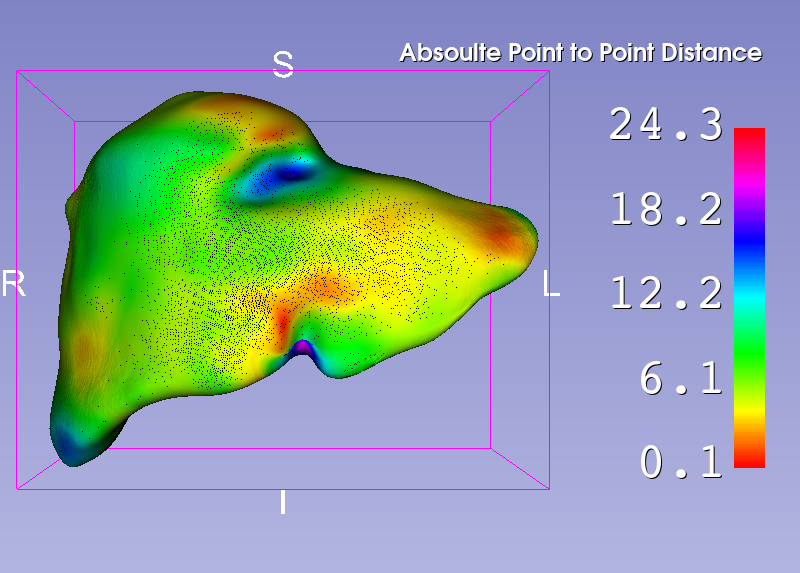}
    %\caption{20\_over}\label{fig:20_over}
  \end{subfigure}
  \begin{subfigure}[b]{0.16\linewidth}
    \includegraphics[width=\linewidth]{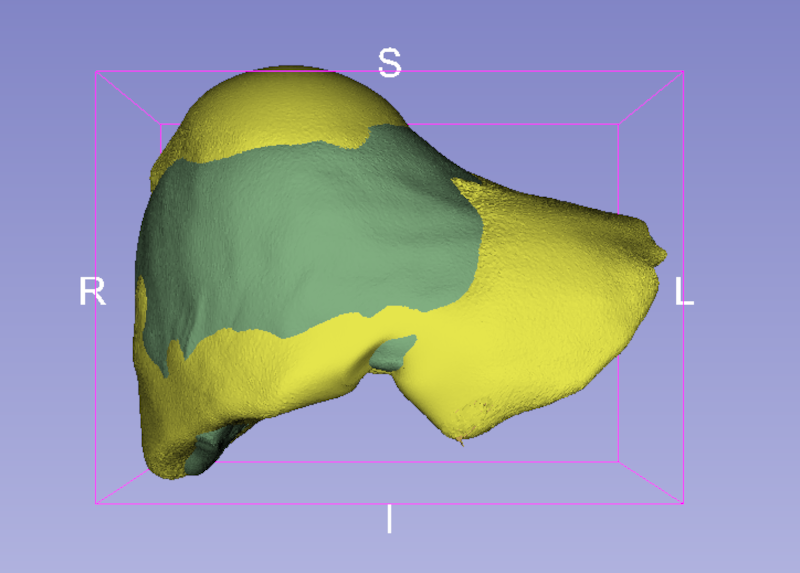}
    %\caption{20\_over}\label{fig:20_over}
  \end{subfigure}
  \begin{subfigure}[b]{0.16\linewidth}
    \includegraphics[width=\linewidth]{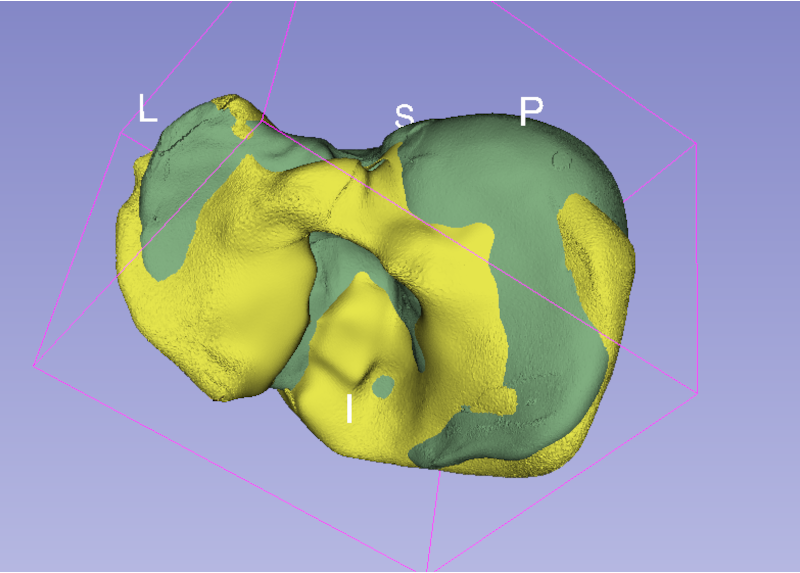}
    %\caption{20\_over\_back}\label{fig:20_over_back}
  \end{subfigure} 
  \begin{subfigure}[b]{0.16\linewidth}
    \includegraphics[width=\linewidth]{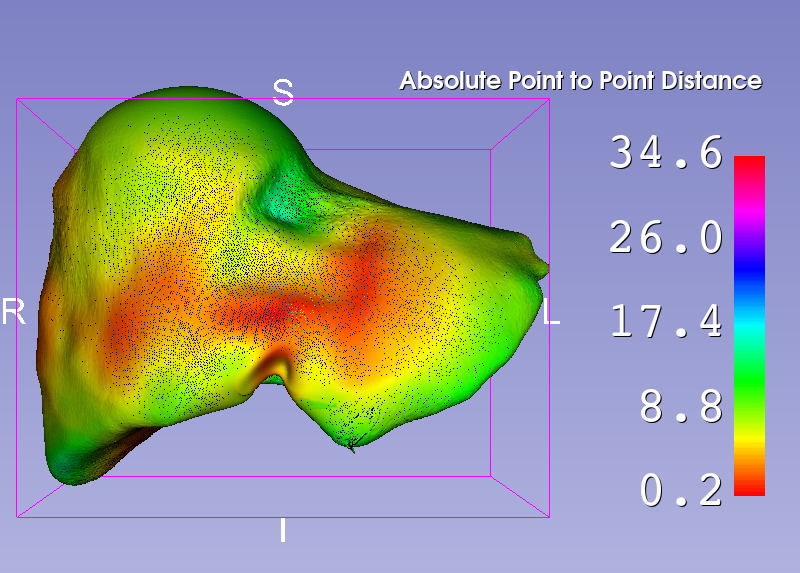}
    %\caption{20\_over\_back}\label{fig:20_over_back}
  \end{subfigure}
  \caption{Three visualizations (anterior, posterior and absolute point to point distance) of reconstruction accuracy of two livers: ground truth (yellow \textcolor[rgb]{0.894,0.894,0.298}{\rule{6pt}{6pt}}, liver models generated from CT segmentation) overlaps well with our results (green \textcolor[rgb]{0.494,0.671,0.494}{\rule{6pt}{6pt}}). }
  \label{fig:3D_reconstructions}
\end{figure}

\noindent{\bf US Liver Segmentation Results:} We used FCN~\cite{fcn}, UNet~\cite{unet}, UNet++~\cite{unet++} with EfficientNetB7 encoder, and TransUNet~\cite{chen2021transunet} for segmenting US scans for US liver segmentation (Table~\ref{ta:accuracy}). TransUNet achieved the best Accuracy (Acc.), Dice Score Coefficient (DSC), Intersection over Union (IoU), and Hausdorff Distance (HD) for unseen data. This is because TransUNet uses transformers to encode tokenized image patches from a CNN feature map. Thus, the input sequence captures global contexts \cite{chen2021transunet}. We used UNet as the decoder to decode the hidden features for generating the final segmentation masks. 2D liver predictions overlap well with ground truth liver labels (Fig.~\ref{fig:overall_flow_results}). This, in turn, leads to an accurate liver volume calculation. Ours is the first method that uses a transformer network in US liver segmentation. Following this result, we used TransUNet for all other experiments. 

\begin{table}[]
\caption{{\bf Main result}: Statistical analysis: RMSE is less in estimated volumes from our method.  Paired $t$-test shows that there is no significant difference in volumes between CT and our method ($p>0.05$). Our method is statistically more accurate. $\mu$: mean difference, SEM: Standard Error of the Mean.}
\label{ta:statistical_analysis}
\centering
\begin{adjustbox}{width=\textwidth}
\begin{tabular}{@{}lr|rrrrrrrr@{}}
\toprule
% \multirow{2}{*}{} & \multirow{2}{*}{{RMSE}}  & \multirow{2}{*}{} & \multirow{2}{*}{{Corr.}} & \multirow{2}{*}{{M}} & \multirow{2}{*}{{SD}} &  \multirow{2}{*}{{SEM}} & \multicolumn{2}{c}{{\begin{tabular}[c]{@{}c@{}}95\% CI of diff.\\ (Lower, Upper)\end{tabular}}} & \multirow{2}{*}{{t}} & \multirow{2}{*}{{df}} &  \multicolumn{2}{r}{{\begin{tabular}[r]{@{}r@{}}Sig.\\ 2-tailed\end{tabular}}} \\ \hline
Vol. Compar. & RMSE & Pair  & $\mu$ & std. & SEM &  \begin{tabular}{r}95\% CI of $\mu$ diff.\\(Lower, Upper)\end{tabular}  & $t$ & df& \begin{tabular}{r}Signi.\\2-tailed\end{tabular}\\
\midrule
CT \& Childs'  &  306.9   &  1   & -201.5 & 234.8 &39.7 & (-282.1, -120.8) & -5.1 & 34 & .000 \\ \midrule
CT \& Ours& {\bf 275.8}&  2  & 78.1 & 268.4 & 45.4 &(-14.1, 170.3) & 1.7 & 34 & .094  \\                         
\bottomrule
\end{tabular}
\end{adjustbox}
\end{table}

\begin{figure}[t]
        \centering
        \input{files/volumes_boxplot}
        \captionsetup{justification=justified}
        \caption{Box plot of liver volumes calculated from Childs' method, CT segmentation, and the proposed method: Childs' method has outliers, but the proposed method has no outliers and its liver volume distribution falls within CT segmentation's liver volume distribution.}
        \label{fig:box_plot}        
\end{figure}
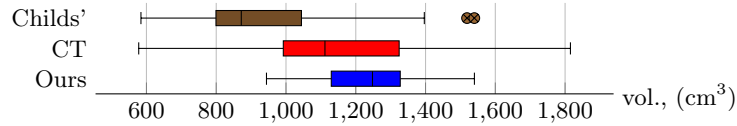

\noindent{\bf 3D Reconstruction Results}: We send the liver masks obtained from the above segmentation process to the 3D reconstruction model to generate the shape parameters to reconstruct the 3D liver shape model. As the problem at hand is a slice mask based shape reconstruction, we use the reconstruction method in Yuan \emph{et al.}~\cite{yuan2022slice}. Table~\ref{ta:accuracy} illustrates the accuracy of the 3D reconstruction method on test data. We use Chamfer Distance (CD) and Mean Surface Distance (MSD) to compare the generated 3D reconstruction with ground truth 3D liver models (see Table~\ref{ta:accuracy}). The combination of TransUNet and parametric regression MLP obtained less CD and MSD compared to using UNet with the same setup. Front and back views of 3D reconstructed liver models are in Fig.~\ref{fig:3D_reconstructions}; our system can generate liver models retaining their complex shape. Further, Fig.~\ref{fig:3D_reconstructions} provides visualization of results, where predicted liver models highly overlap with ground truth. We calculated the Root Mean Square Error (RMSE) to compare with the ground truth CT volumes with the volumes obtained by radiologists using the Childs' method and the proposed method; our liver volumes are closer to the ground truth as shown in Table~\ref{ta:statistical_analysis}.
% We compared the liver volumes calculated using our method and estimates made by radiologists using the Child's method against the CT liver segmentation. 
Box plot in Fig.~\ref{fig:box_plot} shows the descriptive statistics of each liver volume calculation method. We performed patient-wise paired sample $t$-test as shown in Table~\ref{ta:statistical_analysis}. We can conclude that there is a significant difference in liver volumes calculated between CT segmentation ($\mathrm{mean}=1162.4, \mathrm{std.}=275.7$) and Childs' method ($\mathrm{mean}=960.9,\mathrm{std.}=257.9$); $t(34)=-5.1, p=.000$.
%; which has strong correlation ($r=0.6$).%
In contrast, there is no significant difference in liver volumes calculated between CT segmentation ($\mathrm{mean} =1162.4, \mathrm{std.} =275.7$) and our method ($\mathrm{mean} =1240.6, \mathrm{std.} =133.1$); $t(34)=1.7, p=0.094$.%; which has poor correlation ($r=0.3$).

\begin{table}[t]
\caption{Effect of input US image resolution, no. of US slices used as input, no. of principal components used for the SSM, and the no. of CT scans used for the SSM. $\ast$ indicates what we used for the final results. These choices do not affect the final results drastically.}
\label{ta:ablations}
    \begin{subtable}[t]{0.2\linewidth}
        \centering
            % \caption{Our system with input resolution $384 \times 384$   has a small improvement.}
            % \caption{Effect of resolution}
            \label{ta:no_of_sizze}
            \centering
            \begin{tabular}{@{}lr@{}}
            \toprule
            US res.                            & RMSE    \\ \hline
            $192^2$~$\ast$            & 275.83 \\ \hline
            $384^2$             & {\bf 271.63} \\
            \bottomrule
            \end{tabular}
    \end{subtable}
    ~
    \begin{subtable}[t]{0.21\linewidth}
        \centering
        % % \caption{No. of US scans used.}
        \label{ta:no_of_ip}
        \centering
        \begin{tabular}{@{}lr@{}}
        \toprule
        No. slices                           & RMSE   \\ \hline
        2          & 281.65 \\ \hline
        3~$\ast$         & {\bf 275.83} \\
        \bottomrule
        \end{tabular}        
    \end{subtable}
    ~
    \begin{subtable}[t]{0.21\linewidth}
    \label{ta:no_of_pc}
    \centering
    \begin{tabular}{@{}lr@{}}
    \toprule
    No. Comp.                            & RMSE    \\ \hline
    10          & 252.35 \\ \hline
    20          & {\bf 249.90}\\ \hline
    40          & 254.53 \\ \hline
    50~$\ast$  &  275.83 \\ \hline
    70          & 306.82 \\ 
    \bottomrule
    \end{tabular}
    \end{subtable}   
~
    \begin{subtable}[t]{0.24\linewidth}
    % \caption{Correlation :.}
    \label{ta:no_of_ssm}
    \centering
    \begin{tabular}{@{}lr@{}}
    \toprule
    CT scans in SSM                         & RMSE    \\ \hline
    50 (1\textsuperscript{st})         & 291.70 \\ \hline
    50 (2\textsuperscript{nd})          & 285.52 \\ \hline
    60          & 279.96 \\ \hline
    80          & {\bf 275.18} \\ \hline
    100~$\ast$       & 275.83 \\ 
    \bottomrule
    \end{tabular}
    \end{subtable}
\end{table}

\noindent {\bf Ablation Study:} Table~\ref{ta:ablations} shows our ablation studies on the effect of resolution of the three US slices, no. of slices used to compute the shape parameters, the no. of principal components used for SSM, and the no. of CT scans used for SSM. We have chosen to use the resolution of $192\times 192$ to preserve computational resources, 3 slices, 50 principal components, and 100 CT scans for SSM to produce better RMSE. The choices do not drastically affect the final results. 

\section{Conclusion}
We presented an automatic method to obtain the 3D reconstruction of the liver using a few incomplete US scans in the sagittal plane by warping a SSM of the liver. We obtained the SSM as the combination of the mean and principal components of a set of liver meshes obtained from manual segmentation of CT scans. Our reconstruction is accurate, irrespective of the set of CT scans used for the SSM. We carried out the warping of this SSM based on the shape parameters obtained by running the binary segmentation masks of the few incomplete US liver scans through a multi-layer perceptron regressor. Our volumetry results are statistically closer to the ground-truth volumes obtained from CT scans than the volumes computed by radiologists using the Childs’ method. We plan to make our parallel, annotated US and CT scan database available for further research. As future work, we wish to use our volumetry and visualizations for disease diagnosis. 

\begin{credits}
\subsubsection{\ackname} K. Sivayogaraj acknowledges the support received from the Chancellor's scholarship donated by Zone24x7 Incorporated. R. Rodrigo acknowledges the support received from the University of Moratuwa Senate Research Committee grant SRC/LT/2021/20.

\subsubsection{\discintname}
The authors declare that they have no competing interests.
% It is now necessary to declare any competing interests or to specifically
% state that the authors have no competing interests. Please place the
% statement with a bold run-in heading in small font size beneath the
% (optional) acknowledgments\footnote{If EquinOCS, our proceedings submission
% system, is used, then the disclaimer can be provided directly in the system.},
% for example: The authors have no competing interests to declare that are
% relevant to the content of this article. Or: Author A has received research
% grants from Company W. Author B has received a speaker honorarium from
% Company X and owns stock in Company Y. Author C is a member of committee Z.
\end{credits}
\bibliographystyle{splncs04}
\bibliography{LiverUSRecon}




% LNCS Homepage, \url{http://www.springer.com/lncs}, last accessed 2023/10/25
% \end{thebibliography}
\end{document}

%% file: files/main_block_diagram.tex
\def\ehh{0.6}
\begin{tikzpicture}[
    pre/.style={=stealth',semithick},
    post/.style={->,shorten >=1pt,>=stealth',semithick},
    every node/.style={font=\small}
    ]
    \tikzstyle{imagenode} = [inner sep=0pt, outer sep=0pt]
    \coordinate (origin) at (-3,0);
    \def\leftmost{-0.6}
    \draw [fill=rrgreen] (-0.7,\ehh)  -- ++(0.8, -\ehh + 0.1) coordinate (tip) -- ++(0, -0.2) -- ++(-0.8, -\ehh) -- cycle; 
    \draw [fill=rrgreen] (tip) -- ++(0.8, \ehh -0.1) -- ++(0, -1*\ehh) coordinate (tunet) -- ++(0, -1*\ehh)  -- ++(-0.8, \ehh -0.1)  -- ++(0, 0.2) -- cycle; 
    \path (tip) ++(0,-0.8) node[anchor=north, text width=2cm, align=center] {TransUNet segmenter};
    
    \node[imagenode] (raw1) at (\leftmost-1.2,1.1) {\includegraphics[width=1cm]{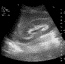}};
    \node[imagenode] (raw2) at (\leftmost-1.2,0) {\includegraphics[width=1cm]{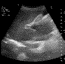}};    
    \node[imagenode] (raw3) at (\leftmost-1.2,-1.1) {\includegraphics[width=1cm]{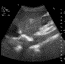}};  
    \draw[post,rounded corners=2pt] (raw1) -- ++(0.7,0) |- ++(0.4, -0.8);
    \draw[post,rounded corners=2pt] (raw2) -- ++(0.7,0) -- ++(0.4, 0);
    \draw[post,rounded corners=2pt] (raw3) -- ++(0.7,0) |- ++(0.4, 0.8); 
    \path (raw3) ++(0, -0.5) node [anchor=north, text width=2cm , align=center]   {US slices};

    \path (tunet) ++(1,1.1) node[imagenode] (mask1)  {\includegraphics[width=1cm]{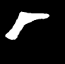}};
    \path (tunet) ++(1,0) node[imagenode]  (mask2) {\includegraphics[width=1cm]{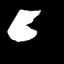}};    
    \path (tunet) ++(1,-1.1)  node[imagenode]  (mask3) {\includegraphics[width=1cm]{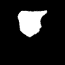}}; 

    \path (mask3) ++(0, -0.5) node [anchor=north, text width=2cm, align=center]   {Masks};
    \draw[post,rounded corners=2pt] (tunet) ++(0,0.3) -- ++(0.2,0)  |- (mask1);
    \draw[post,rounded corners=2pt] (tunet) ++(0,0) -- ++(0.2,0)  -- (mask2);
    \draw[post,rounded corners=2pt] (tunet) ++(0,-0.3) -- ++(0.2,0)  |- (mask3);

    \path (mask2) ++(2.2,0) node[fill=rrteal, draw=black, rounded corners=2pt, text width=1.5cm, align=center] (mlp) {Param. regress. MLP};

    \draw (mask2) ++(0.9, 0) node[draw=black, inner sep=2pt] (s) {S};
    \draw[post,rounded corners=2pt] (mask1) -| (s);
    \draw[post,rounded corners=2pt] (mask2) -- (s);
    \draw[post,rounded corners=2pt] (mask3) -| (s);
    \draw[-stealth] (s) -- (mlp);
    \path (s) ++ (-0.1, -1.5) node[anchor=west] {S: stacking};

    \draw[-stealth] (mlp) -- ++(1.5,0)  node[fill=rrteal, draw=black, text width=3cm, align=center, rotate=90] (sp) {Shape param. $\{\alpha_k\}$};
    \draw[-stealth] (mlp) -- (sp);
    \path (sp) -- ++(1.3,0) node[imagenode] (3d) {\includegraphics[width=1.2cm]{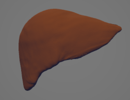}};
    \path (3d) ++(0, -0.8) node [anchor=north, text width=2cm, align=center]   {3D reconstruction};
    \draw[-stealth] (sp) -- (3d);
    \path (3d) ++(0,1.2) node[fill=rrlthellow, draw=black, rounded corners=2pt, text width=1.1cm, align=center] (ssm) {SSM $\bar{s},\{v_k\}$};
    \draw[-stealth] (ssm) -- (3d);
    \draw[-stealth] (3d) -- ++(1,0) node[anchor=west,draw=black, rounded corners=2pt, text width=0.9cm, align=center] {Vol. $\mathrm{cm}^3$};
\end{tikzpicture}

%% file: files/volumes_boxplot.tex
\begin{tikzpicture}
	\pgfplotstableread[col sep=comma]{files/volumes.csv}\csvdata
	% Boxplot groups columns, but we want rows
	% \pgfplotstabletranspose\datatransposed{\csvdata} 
	\begin{axis}[
        y = 0.4cm,
		boxplot/draw direction = x,
		y axis line style = {opacity=0},
		axis x line* = bottom,
		% enlarge x limits,
        xlabel = {vol., ($\mathrm{cm}^3$)},
        x label style={at={(axis description cs:1.0,0.4)},anchor=west},  
		xmajorgrids,
		xticklabel style = {align=center, font=\small},
  	ytick = {1, 2, 3},
		yticklabels = {Ours, CT, Childs'},
		ytick style = {draw=none}, % Hide tick line
        boxplot/box extend = 0.5,
	]
		\foreach \n in {1,...,3} {
			\addplot+[boxplot, fill, draw=black] table[y index=\n] {\csvdata};
		}
	\end{axis}
\end{tikzpicture}